# Time-resolved shadowgraphs of transient plasma induced by spatiotemporally focused femtosecond laser pulses in fused silica glass

Zhaohui Wang,[1,2] Bin Zeng,[1] Guihua Li,[1] Hongqiang Xie,[1,2] Wei Chu,[1] Fei He,[1] Yang Liao,[1] Weiwei Liu,[3] Hui Gao,[4] and Ya Cheng[1,5*]

[1] *State Key Laboratory of High Field Laser Physics, Shanghai Institute of Optics and Fine Mechanics, Chinese Academy of Sciences, Shanghai 201800, China*
[2] *University of Chinese Academy of Sciences, Beijing 100049, China*
[3] *Institute of Modern Optics, Nankai University, Key Laboratory of Optical Information Science and Technology, Ministry of Education, Tianjin 300071, China*
[4] *School of Science, Tianjin Polytechnic University, Tianjin 300387, China*
[5] *Collaborative Innovation Center of Extreme Optics, Shanxi University, Taiyuan, Shanxi 030006, People's Republic of China*
*Corresponding author: ya.cheng@siom.ac.cn

**We report on experimental observations of formation and evolution of transient plasma produced in fused silica glass with spatiotemporally focused (STF) femtosecond laser pulses using a pump-probe shadow imaging technique. Surprisingly, the observation shows that the track of the plasma is significantly curved, which is attributed to an asymmetric density distribution of the transient plasma produced in the focal volume caused by the pulse front tilt of the STF laser field.**

Femtosecond laser three-demensional (3D) micromachining has attracted much attention due to a wide range of applications in many research areas such as microfluidics, micro-optics, micro-electronics and optofluidics [1-8]. Once the femtosecond laser pulses are tightly focused in transparent materials, nonlinear interactions such as multiphoton absorption and avalanche ionization are efficiently confined within the focal volume, leading to the generation of a dense electron plasma. Such transient plasma may induce a permanent micro- or nanoscale internal modification of bulk transparent materials, which provides the unique capability of 3D micro- and nano fabrication [9-11]. Unlike the conventional two-dimensional (2D) planar lithographic fabrication in which only the transverse fabrication resolution should be considered, the longitudinal (i.e., the axial direction which is parallel to the propagation direction of the laser beam) fabrication resolution plays an equally important role in 3D femtosecond laser processing in terms of the quality of the fabricated structures. For many practical applications such as writing optical waveguides or fabricating microchannels, a 3D isotropic resolution is necessary. Recently, it has been reported that the spatiotemporal beam shaping technique allows 3D isotropic resolution to be achieved with a single objective lens [12]. The trick of this technique is to separate the spectral components of a femtosecond pulse spatially and recombine them using an objective lens. The spatial overlap of different frequency components only happens at the focal volume, thus the pulse width is the shortest only at the focal volume while the pulse width out of focus is stretched. The peak intensity is strongly localized along the axis in both space and time domains, and consequently, the longitudinal resolution is improved.

Interestingly, it was discovered that the spatiotemporally focused spot has some unconventional characteristics in addition to the varying pulse width during propagation, including a tilted pulse front [13] and a tilted peak intensity distribution [14]. In particular, the tilted pulse front can induce a nonreciprocal writing effect which gives rise to an anisotropic fabrication quality depending on the direction of sample translation inside an isotropic medium. Until now, the mechanism behind the nonreciprocal writing effect has not been fully understood, although the intimate connection of this effect to the plasma dynamics controlled by the pulse front tilt (PFT) of femtosecond pulses has been confirmed by several recent investigations [13,15]. Here, to look into the insight of the nonlinear interaction of spatiotemporally focused pulses with transparent materials, we employ a pump-probe shadow imaging technique to observe the formation and evolution of the transient plasma induced by the STF pulses. We compare the results obtained by spatiotemporal focusing (STF) and conventional focusing (CF) systems to reveal the different plasma behaviors of the two focusing conditions.

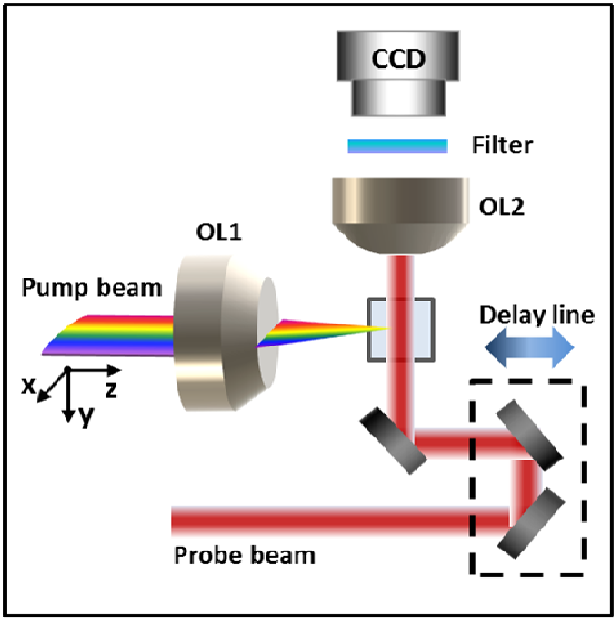

Fig. 1. Schematic of the experimental setup.

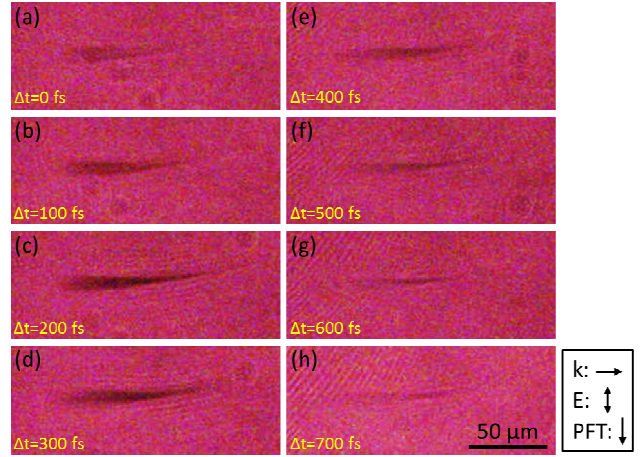

Fig. 2. Time-resolved shadowgraphs of transient plasma induced by the STF femtosecond pulse inside fused silica glass when the polarization direction of the pump pulse is parallel to the spatial chirp. The time-delay is directly indicated in each panel. The directions of the laser propagation, polarization of the laser pulse, and the PFT are indicated by the arrows next to the image.

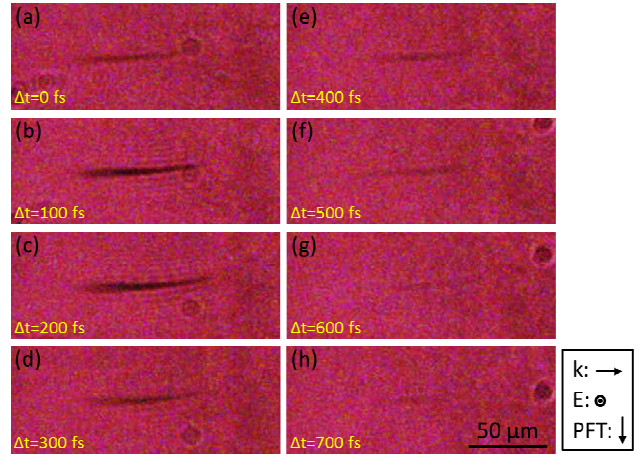

Fig. 3. Time-resolved shadowgraphs of transient plasma induced by the STF femtosecond pulse inside fused silica glass when the polarization direction of the pump pulse is perpendicular to the spatial chirp. The time-delay is directly indicated in each panel. The directions of the laser propagation, polarization of the laser pulse, and the PFT are indicated by the arrows next to the image.

Figure 1 schematically illustrates the experimental setup. The femtosecond laser system (Legend-Elite, Coherent Inc.) consists of a Ti:sapphire laser oscillator and amplifier, and a grating-based stretcher and compressor, which delivers 4 mJ, 50 fs pulses with a spectral bandwidth of ~26 nm centered at 800 nm wavelength at 1 kHz repetition rate. During the experiment, the femtosecond laser system operated in the single shot mode. Before the amplified laser beam was recompressed, it was split into two using a 1:1 beam splitter. One beam, which was used to produce the spatiotemporally focused pump beam, was spatially dispersed along the $x$ direction by a pair of 1500 lines/mm gratings (blazing at ~53°). More details on the arrangement of the gratings can be found elsewhere [14].

A half-wave plate and a variable neutral density filter were inserted in series before the gratings to adjust the power and the polarization of the pump beam. After being dispersed by the grating pair, the beam size was reduced to ~10 mm ($1/e^2$) along the $x$-axis and 2 mm ($1/e^2$) along the $y$-axis using a telescope system consisting of a convex lens ($f$ = 50 cm) and a concave lens ($f$ = -10 cm). The beam was then focused into a fused silica glass sample at a depth of ~250 μm below the surface using an objective lens (20×, NA = 0.40). The glass samples have a size of 5 mm × 5 mm × 1 mm, and are polished on all the six sides. The other beam was compressed to 50 fs pulses by passing through the compressor to generate the probe beam. A delay line was inserted into the optical path of the probe beam to adjust the relative time delay between the two beams. The interaction region in the sample was illuminated by the probe beam and then imaged with an objective lens (10×, NA = 0.30) on a CCD camera. Note that the propagation direction of the probe beam was always perpendicular to the plane of the spatial chirp of the pump beam as shown in Fig. 1. A bandpass filter centered at 800 nm (40 nm bandwith) was inserted in front of the CCD camera to remove the plasma emission contribution to the transmitted probe signal. For the single-shot measurement required by our investigation, the sample was shifted to a fresh area after each shot of the laser pulse. Since we are more interested in observation of spatial property of the plasma generated inside glass rather than the plasma dynamics, we define the zero time delay (Δt = 0 fs) as the moment the transient plasma appeared in the field of view (i.e., for the zero time delay defined here, the pump and probe pulses do not necessarily overlap in time).

Figure 2 shows a series of shadowgraphs of the transient plasma induced by the spatiotemporally focused pulse recorded at different time delays. The polarization direction of the pump laser pulse was set along $x$ axis (see, Fig. 1) which was parallel to the spatial chirp. The pump pulse was incident from the left, and the pulse energy was set to be 6 μJ. The direction of the pulse front tilt, which indicates the sweeping direction of the pulse front in the focal volume, is from the top side to the bottom as indicated in Fig. 2, and the value of the PFT in the focal plane is calculated to be 50 fs/μm based on our experimental parameters. It should be noted that such a large PFT usually does not exist in conventional femtosecond laser pulses even if the laser system is not perfectly aligned. The dark region in these shadowgraphs results from the absorption of the probe beam by the laser-induced electron plasma inside the glass. When the time delay is short (Δt < 200 fs), the absorption of the probe beam increases with the increasing time delay. At the time delay of Δt = 200 fs, the absorption was maximized, which indicates that the density of the plasma reached the peak. At the longer time delays, the plasma started to decay due to the recombination of the electrons. Remarkably, it can be observed that the track of the plasma was bent during its propagation. A bending angle of ~8° was formed between the track of the plasma and the optical axis of the lens. When the polarization direction of fundamental laser was switched to be parallel to the *y* axis, which is perpendicular to the spatial chirp, the similar phenomenon of the plasma dynamics were observed as shown in Fig. 3. Nevertheless, by making a comparison between the results in Fig. 2 and Fig. 3, we find that when the polarization direction of the pump pulse is parallel to the spatial chirp, the plasma formation appears more pronounced. Thus, stronger photoionization has occurred in such a condition. We will discuss this feature later.

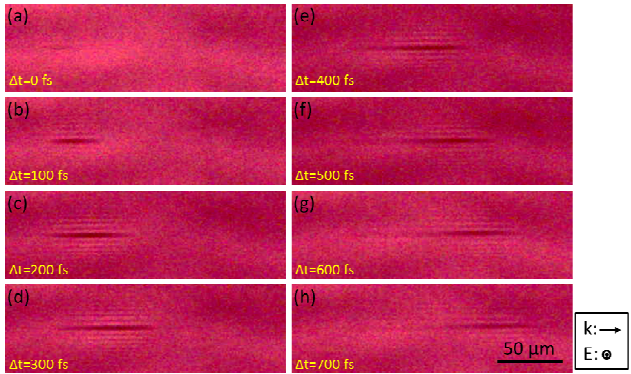

Fig. 4. Time-resolved shadowgraphs of transient plasma induced by the CF femtosecond pulse inside fused silica glass. The time-delay is directly indicated in each panel. The propagation and the polarization directions of the pump laser pulse are indicated by the black arrows next to the image.

For comparison, we performed the same set of experiments using a CF system, which was achieved by simply replacing the grating pair with a pair of flat gold mirrors. Since the CF system produces a longer plasma track due to the low numerical aperture and the removal of the temporal focusing effect, the pulse energy was decreased to 2.5 μJ to avoid the ionization on the surface of the sample. As shown in Fig. 4, at the smallest time delay of Δt = 0 fs, a weak plasma was formed by the laser induced photoionization. With the increasing time delay, the plasma channel extends its length along the propagation direction of the pump pulse, which can be understood as a combined effect of the propagation of the laser pulses in the self-focused region and the survival of the laser produced plasma within its limited lifetime. At a time delay of Δt = 300 fs, the plasma channel formed in the early stage during the propagation of the pump pulses (e.g., the plasma appeared closer to the front surface as shown in Fig. 4(a)) started to decay. Meanwhile, the plasma channel kept extending its length along the propagation direction, resulting in a total plasma length of ~180 μm as shown in Fig. 4(h), which is much longer compared with the plasma lengths obtained by the STF system in Fig. 2 and Fig. 3. Moreover, unlike the results obtained with the STF system, the plasma tracks appear perfectly straight in Fig. 4 due to the symmetrical geometry of the Gaussian-like focal spot produced in the glass sample by the CF system.

The curved tracks of plasma distribution in Fig. 2 and Fig. 3, which imply that the light is propagating along a curved trajectory, has not been reported before in a STF geometry. It should be noted that the phenomenon observed in the experiment is fundamentally different from the intensity plane tilt (IPT) of the focal spot previously observed in linear or weak nonlinear propagation conditions [14]. Using the theoretical tool in the previous work [14], the calculated confocal length of the focal spot produced in fused silica glass with the STF system is ~8 μm by assuming the linear propagation condition, which is one order of magnitude shorter than the length of plasma tracks observed in the experiment (e.g., ~80 μm as shown in Fig. 3). In another word, the IPT in the focal spot only occurs in a highly localized region near the geometric focal plane of the lens, which cannot contribute to the formation of curved plasma channels over the propagation distances significantly longer than the Rayleigh length. To form the curved plasma channels, the center of the focal spot where the strongest photoionization occurs must be able to propagate along a curve trajectory.

To gain a deeper insight into such phenomenon, we further examined the plasma distribution as we varied the polarization directions as well as the pulse energy. The results are shown in Fig. 5. In this observation, the time delay was fixed at Δt = 200 fs. The plasma tracks are all bent along the propagation of the pump laser pulses, whereas the bending angles are not the same for the different laser conditions. The angles measured in Figs. 5 (a) - (f) are ~8°, ~8°, ~5°, ~8°, ~5°, and ~4°, respectively. The results suggest that the plasma track bends itself more severely when the photoionization appears stronger from the shadowgraph images. Thus, some nonlinear optical effects should be taken into account to understand the mechanism behind such phenomenon.

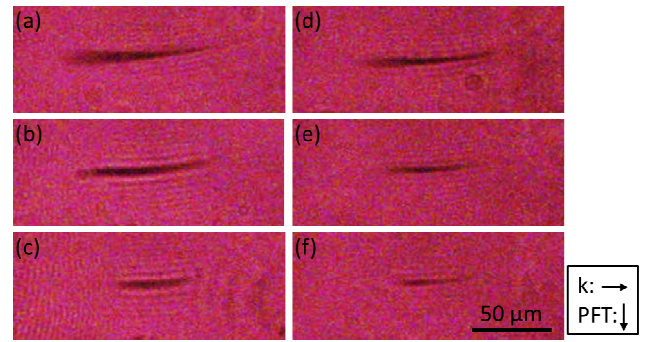

Fig. 5. Time-resolved shadowgraphs of transient plasma induced by spatiotemporally focused femtosecond pulses inside fused silica glass. The directions of the laser propagation and the PFT are both indicated by the black arrows next to the image. The polarization direction of the femtosecond laser pulse is parallel to the spatial chirp in (a)-(c), but it is perpendicular to the spatial chirp in (d)-(f). The pulse energies are 6 μJ in (a), (d), 4 μJ in (b),(e) and 2.5 μJ in (c), (f), respectively.

Based on the strong dependence of the bending angle on the peak intensity and the polarization direction of the femtosecond laser pulses, we tentatively attribute the bending of the plasma track to an asymmetric plasma expansion during the interaction of the STF pulses with the fused silica glass. Here, the PFT in the focal plane of the STF spot reaches 50 fs/μm, i.e., the intensity front of the pulse rapidly sweeps across the focal plane at a velocity of $2\times10^7$ m/s, which is only about one order of magnitude smaller than the phase velocity of light in fused silica. Therefore, in the cross section of the self-focused femtosecond laser beam (i.e., along the transverse plane perpendicular to the optical axis of the focal lens) produced by the STF system, the photoionization always starts from one side and sweeps across the cross section along the direction of the PFT. The rapidly moving ionization front would force the plasma to expand asymmetrically, i.e., the plasma expansion is more pronounced along a direction of the PFT than along the opposite direction. The asymmetric expansion further leads to the formation of an inhomogeneous distribution of the plasma density in the cross section of the plasma tracks.

The asymmetrical plasma density distribution will give rise to an inhomogeneous distribution of the refractive index in the focal volume. In this case, the refractive index would decrease more in the region of high-density plasma. This effect is equivalent to bending the light with a prism, as the STF pulse produces a transient wedge of plasma. Previously, it has shown that the plasma density produced by the irradiation of the femtosecond laser pulses in fused silica can be on the order of $10^{19}/cm^3$, corresponding to a refractive index change as high as 0.01 [16]. The gradient of the refractive index induced by such high-density plasma is sufficient to bend the propagation trajectory of the femotosecond pulses. Specifically, we notice that the bending direction of the plasma tracks is consistent with this scenario, because the light travelling through the plasma wedge will bend toward the region of lower plasma density (i.e., the region of higher refractive index). Furthermore, when the photoionization is stronger with the higher pulse energies used, a larger gradient of the refractive index will be created because of the enhanced asymmetry in the plasma density distribution. In this case, larger bending angles should be observed, which is exactly what we have seen in Fig. 5. When the CF scheme is chosen, the distribution of the plasma density always has an angular symmetry due to the Gaussian-like spatial profile of the femtosecond laser pulses. Thus only the plasma defocusing occurs which is well known in the nonlinear propagation regime. The propagation direction of the laser beam will always be along the optical axis of the focal lens, forming the straight plasma tracks as shown in Fig. 4.

Another interesting observation is that the photoionization appears stronger when the polarization direction of the pump laser is parallel to the spatial chirp as shown by the results in Fig. 2, Fig. 3 and Fig. 5. It is noteworthy that after being ionized in the strong laser fields, the free electrons quiver back and forth along the direction of the polarization of the laser beam. After an average of the quiver motion over the optical cycle, the free electrons experience a ponderomotive force in the inhomogeneous light field, which is expressed as

$$F_p = -\frac{e^2}{4m_e\omega^2}\nabla I , \qquad (1)$$

where $e$ is the electron charge, $m_e$ is the electron mass, $\omega$ is the frequency of the light and $I$ is the light intensity [17,18]. The direction of the ponderomotive force is also parallel to the polarization direction of the laser beam. When the polarization of the laser is parallel to the direction of the spatial chirp, the direction of the pondermotive force would be parallel to the direction of the PFT, making a constructive contribution to the plasma expansion. This could lead to stronger plasma due to the enhanced avalanche ionization, as the avalanche ionization sensitively depends on the density of hot electrons. The constructive contribution disappears when the polarization direction of the pump laser is set to be perpendicular to the direction of the spatial chirp, as the ponderomotive force, which is perpendicular to the PFT under such circumstance, will no longer be able to enhance the plasma expansion.

It is worth mentioning that previous numerical investigations on the nonlinear propagation of the spatiotemporally focused pulse in transparent media do not reveal the bending of the propagation trajectory of the laser pulse because both the plasma expansion and the ponderomotive force are neglected in the simulations [19,20]. Our results imply that these plasma effects play a significant role that cannot be ignored in the nonlinear propagation of intense STP pulses. Modeling tools involving both the effects need to be developed in the future for quantitatively understanding the interaction of intense STF pulses with transparent materials.

In summary, we have investigated behaviors of the transient plasma induced by the intense STF pulse inside fused silica glass using a pump-probe shadow imaging technique. Curved tracks of plasma distribution are observed which is unique with the STF scheme. The phenomenon can be interpreted based on the inhomogeneous plasma distributions produced in the intense STF laser field. A strong PFT introduced by the spatial chirp of the incident femtosecond pulse is responsible for producing the asymmetrical plasma expansion in the focal volume. We stress that the photoionization is the most fundamental process in inducing various types of modifications in transparent materials with femtosecond laser pulses, thus the observed bending in the plasma tracks is crucial for understanding the exotic phenomena in centrosymmetric solid or gaseous media such as nonreciprocal writing and second harmonic generation [15,18]. Our finding will have implications in various applications ranging from micromachining, filamentation optics to remote atmospheric sensing.

**References**

1. R. R. Gattass and E. Mazur, "Femtosecond laser micromachining in transparent materials," Nature Photon. **2**, 219 (2008).
2. K. Sugioka and Y. Cheng, "Ultrafast lasers—reliable tools for advanced materials processing," Light: Sci. Appl. **3**, e149 (2014).
3. R. Osellame, H. J. Hoekstra, G. Cerullo, and M. Pollnau, "Femtosecond laser microstructuring: an enabling tool for optofluidic lab‐on‐chips," Laser Photon. Rev. **5**, 442 (2011).
4. M. Beresna, M. Gecevičius, and P. G. Kazansky, "Ultrafast laser direct writing and nanostructuring in transparent materials," Adv. Opt. Photon. **6**, 293 (2014).
5. K. Itoh, W. Watanabe, S. Nolte, and C. Schaffer, "Ultrafast processes for bulk modification of transparent materials," MRS Bull. **31**, 620 (2006).
6. M. Ams, G. D. Marshall, P. Dekker, J. A. Piper, and M. J. Withford, "Ultrafast laser written active devices," Laser Photon. Rev. **3**, 535 (2009).
7. F. Chen and J. R. Aldana, "Optical waveguides in crystalline dielectric materials produced by femtosecond‐laser micromachining," Laser Photon. Rev. **8,** 251 (2014).
8. K. Sugioka and Y. Cheng. "Femtosecond laser processing for optofluidic fabrication," Lab Chip, **12**, 3576 (2012).
9. Y. Bellouard, A. Champion, B. Lenssen, M. Matteucci, A. Schaap, M. Beresna, C. Corbari, M. Gecevicius, P. Kazansky, O. Chappuis, M. Kral, R. Clavel, F. Barrot, J. M. Breguet, Y. Mabillard, S. Bottinelli, M. Hopper, C. Hoenninger, E. Mottay, and J. Lopez, "The Femtoprint Project," J. Laser Micro/Nanoeng. **7**, 1 (2012).
10. S. Juodkazis, K. Nishimura, S. Tanaka, H. Misawa, E. G. Gamaly, B. Luther-Davies, L. Hallo, P. Nicolai, and V. T. Tikhonchuk, "Laser-induced microexplosion confined in the bulk of a sapphire crystal: evidence of multimegabar pressures," Phys. Rev. Lett. **96**, 166101 (2006).
11. Y. Liao, Y. Cheng, C. Liu, J. Song, F. He, Y. Shen, D. Chen, Z. Xu, Z. Fan, X. Wei, K. Sugioka, and K. Midorikawa, "Direct laser writing of sub-50 nm nanofluidic channels buried in glass for three-dimensional micro-nanofluidic integration," Lab Chip **13**, 1626 (2013).
12. F. He, H. Xu, Y. Cheng, J. Ni, H. Xiong, Z. Xu, K. Sugioka, and K. Midorikawa, "Fabrication of microfluidic channels with a circular cross section using spatiotemporally focused femtosecond laser pulses," Opt. Lett. **35**, 1106 (2010).
13. D. N. Vitek, E. Block, Y. Bellouard, D. E. Adams, S. Backus, D. Kleinfeld, C. G. Durfee, and J. A. Squier, "Spatio-temporally focused femtosecond laser pulses for nonreciprocal writing in optically transparent materials," Opt. Express **18**, 24673 (2010).
14. F. He, B. Zeng, W. Chu, J. Ni, K. Sugioka, Y. Cheng, and C. G. Durfee, "Characterization and control of peak intensity distribution at the focus of a spatiotemporally focused femtosecond laser beam," Opt. Express **22**, 9734 (2014).
15. P. G. Kazansky, W. Yang, E. Bricchi, J. Bovatsek, A. Arai, Y. Shimotsuma, K. Miura, and K. Hirao, "Quill writing with ultrashort light pulses in transparent materials," Appl. Phys. Lett. **90**, 151120(2007).

16. Q. Sun, H. Jiang, Y. Liu, Z. Wu, H. Yang, and Q. Gong, " Measurement of the collision time of dense electronic plasma induced by a femtosecond laser in fused silica," Opt. lett. **30**, 320(2005).
17. G. Li, L. Ni, H. Xie, B. Zeng, J. Yao, W. Chu, H. Zhang, C. Jing, F. He, H. Xu, Y. Cheng, and Z. Xu, " Second harmonic generation in centrosymmetric gas with spatiotemporally focused intense femtosecond laser pulses," Opt. lett. **39,** 961 (2014).
18. M. Beresna, P. G. Kazansky, Y. Svirko, M. Barkauskas, and R. Danielius, "High average power second harmonic generation in air," Appl. Phys. Lett. **95**, 121502 (2009).
19. Zeng, W. Chu, H. Gao, W. Liu, G. Li, H. Zhang, J. Yao, J. Ni, S, Chin, Y. Cheng, and Z. Xu, "Enhancement of peak intensity in a filament core with spatiotemporally focused femtosecond laser pulses," Phys. Rev. A **84**, 063819(2011).
20. T. Xi, Z. Zhao, and Z. Hao, " Filamentation of femtosecond laser pulses with spatial chirp in air," J. Opt. Soc. Am. B **31**, 321 (2014).